\documentclass[a4paper]{article}
\usepackage{graphicx}
\usepackage{amsmath}
\usepackage{url}
\usepackage{color}
\usepackage{titlesec}

\def\inststor{Institute}
\def\emailstor{Email}
\def\datestor{\today}
\def\abststor{Abstract}
\def\kwordstor{Keywords}

\newcommand{\inst}[1]{\def\inststor{#1}}
\newcommand{\email}[1]{\def\emailstor{#1}}
\newcommand{\recdate}[1]{\def\datestor{#1}}
\newcommand{\abst}[1]{\def\abststor{#1}}
\newcommand{\kword}[1]{\def\kwordstor{#1}}

\titleformat*{\section}{\large\bfseries}

\date{
	{\small \inststor\\
		\href{mailto:\emailstor}{\emailstor}}\\
	\vspace{10pt}
	\abstract{\abststor}
	}

\usepackage[dvipdfm]{hyperref}
\usepackage{txfonts} 
\usepackage{upgreek}
\usepackage{bm}

\newcommand{\dif}[1]{\mathrm{d} #1 \,}
\newcommand{\mr}[1]{\mathrm{#1}}
\newcommand{\kT}{\vec{k_\mr{T}}}
\newcommand{\kTs}{k_\mr{T}}
\newcommand{\pT}{\vec{p_\perp}}

\newcommand{\phiH}{\varphi_\mr{h}}
\newcommand{\PhT}{\vec{P_\mr{T}}}
\newcommand{\PhTs}{P_\mr{T}}
\newcommand{\A}[2]{A_\mr{#1}^{#2}}
\renewcommand{\vec}[1]{\bm{#1}}

\title{Azimuthal Asymmetries in Unpolarised Semi-Inclusive DIS at COMPASS}

\author{Jan \textsc{Matousek}$^{1}$ on behalf of the COMPASS Collaboration}

\inst{$^{1}$Faculty of Mathematics and Physics,
		Charles University, Ke Karlovu 3, 150\,00 Praha, Czechia}

\email{jan.matousek@cern.ch}

\recdate{January 14, 2022}

\abst{In semi-inclusive deep inelastic scattering (SIDIS) the non-zero transverse momentum of partons induces azimuthal dependence of the cross-section. For an unpolarised nucleon, three azimuthal modulations that can be related to different combinations of twist-two or higher-twist transverse momentum dependent PDFs and fragmentation functions arise: the so-called Cahn effect reflected in a $\cos\phiH$ modulation, the $\cos2\phiH$ term related to the Boer--Mulders PDF and $\sin\phiH$ effect known as beam-spin asymmetry.
In 2016 and 2017, the COMPASS experiment at CERN collected a large sample of SIDIS events using a longitudinally polarised 160~GeV/$c$ muon beam scattering on a liquid hydrogen target. Amplitudes of the aforementioned azimuthal modulations have been extracted from part of the data. A new procedure has been developed to subtract a background coming from the decay of diffractively produced vector mesons. The results presented in this talk qualitatively agree with earlier COMPASS results obtained with an isoscalar target.
}

\kword{DIS, SIDIS, asymmetry, Cahn, Boer--Mulders, COMPASS}

\begin{document}
\maketitle

\section{Introduction}
\begin{figure}[tbh]
\begin{minipage}[b]{0.3\textwidth}
\includegraphics[width=\textwidth]{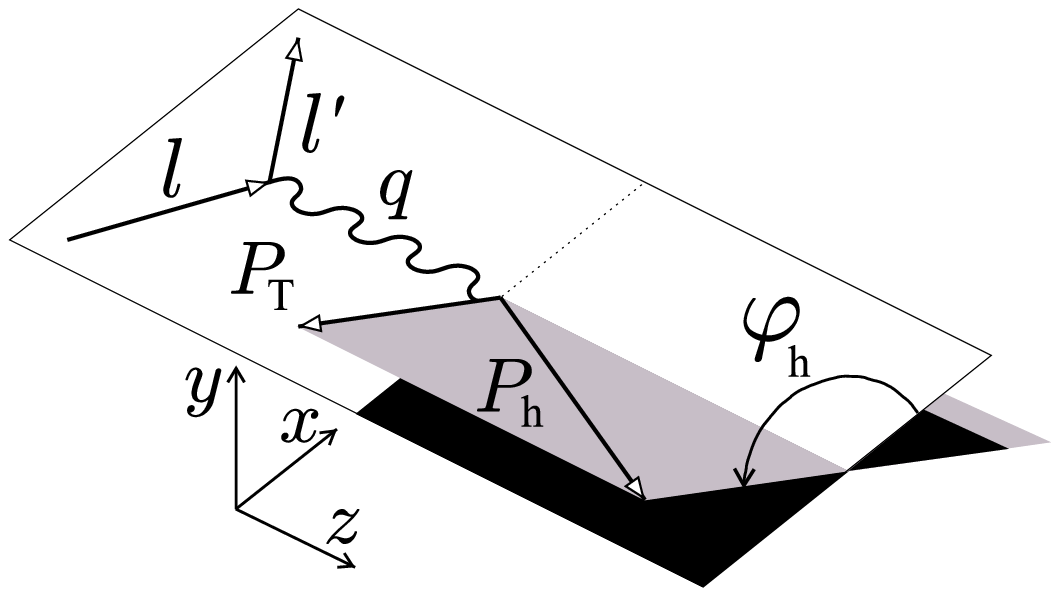}
\caption{\label{fig:gammaN}
	The momentum $\vec{q}$ of the $\upgamma^*$
	defines the $z$ axis of the $\upgamma^*$N system. $\vec{P_\mr{h}}$ is the momentum
	of the observed hadron h.}
\end{minipage}
\hspace{0.01\textwidth}
\begin{minipage}[b]{0.68\textwidth}
\includegraphics[width=\textwidth]{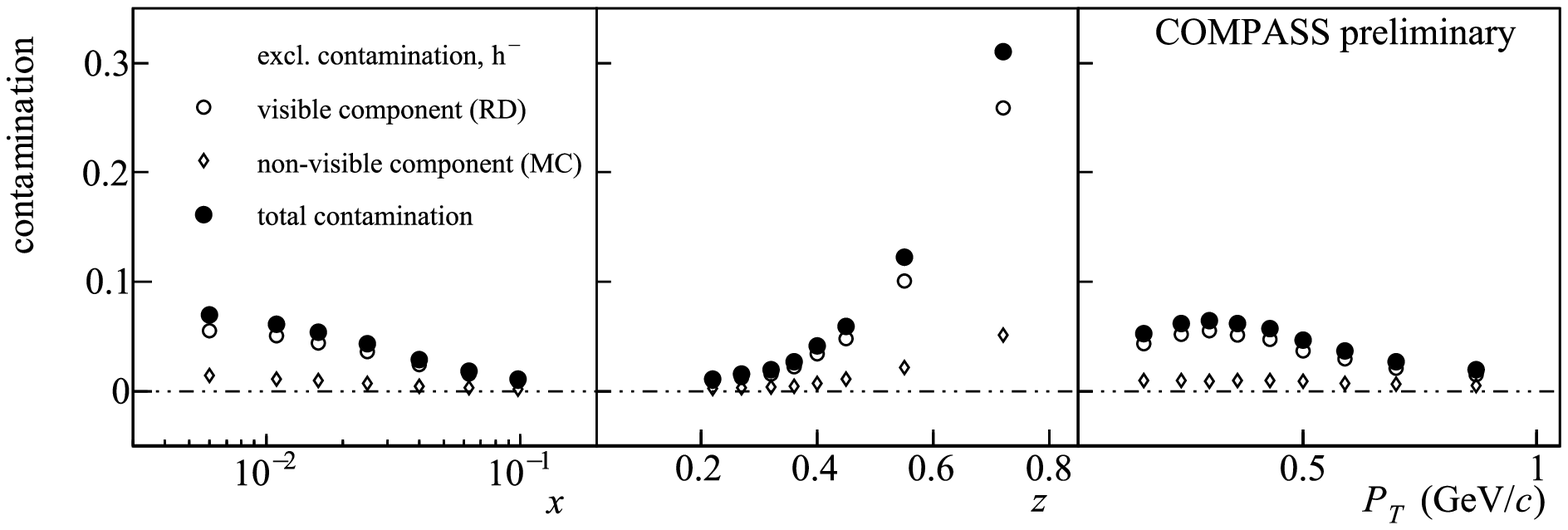}
\caption{\label{fig:cont}
	Fraction of h$^-$ from the decay of exclusive vector mesons in the
	hadron sample before the dedicated selection cut to remove the visible
	component and before the Monte Carlo driven subtraction
	of the non-visible component. The fraction of h$^+$ is slightly lower
	and similar in shape.}
\end{minipage}
\end{figure}

In QCD the momentum and spin structure of the nucleon explored in hard-scattering reactions beyond the collinear approximation is described in terms of transverse momentum dependent parton distribution functions (TMD PDFs). The production of hadrons in the semi-inclusive deep inelastic scattering (SIDIS) of leptons off a nucleon,
$
	\ell \mr{N} \rightarrow \ell^\prime \mr{h X}
$
is a standard tool to probe the TMD nucleon structure. The differential cross-section for an unpolarised nucleon in the one-photon exchange approximation~\cite{Bacchetta:2006tn} can be written as
\begin{align}
	\label{eq:xsec}
	\frac{\dif\sigma}{\dif{x}\dif{y}\dif{z}\dif{\PhTs^2}\dif{\phiH}}
    = \sigma_0\left( 1 + \epsilon_1 \A{UU}{\cos\phiH} \cos\phiH
    	+ \epsilon_2 \A{UU}{\cos2\phiH} \cos2\phiH
        + \lambda \epsilon_3 \A{LU}{\sin\phiH} \sin\phiH \right),
	\\
	\mr{where}
	\quad
	\epsilon_1 = \frac{2(2-y)\sqrt{1-y}}{1+(1-y)^2},
	\qquad
	\epsilon_2 = \frac{2(1-y)}{1+(1-y)^2},
	\qquad
	\epsilon_3 = \frac{2y\sqrt{1-y}}{1+(1-y)^2}.
\end{align}
Here $\PhT=(\PhTs\cos\phiH,\PhTs\sin\phiH)$ is the transverse momentum of the hadron h in the $\upgamma^*$N system outlined in Fig.~\ref{fig:gammaN}, $z$ is the fraction of the photon energy carried by the hadron, $y$ and Bjorken $x$ are the usual DIS variables, $\sigma_0$ is the $\phiH$-independent cross-section, $\lambda$ is the beam longitudinal polarisation and the amplitudes $A_X^Y$ of the azimuthal modulations, also called asymmetries, carry information on the internal structure of the nucleon and on quark fragmentation. Another valuable source of information, which is not covered here, is the $\PhTs^2$-dependence of the $\sigma_0$.

In the framework of the TMD factorisation and up to the order $1/Q$ (twist-three), the amplitudes can be interpreted as convolutions of TMD PDFs and TMD fragmentation functions (TMD FFs) over the intrinsic quark transverse momentum $\kT$ and the transverse momentum $\pT$ obtained in the fragmentation~\cite{Bacchetta:2006tn}:
\begin{equation}\label{eq:tmds}
	\A{UU}{\cos\phiH} = \frac{2M_\mr{p}}{Q} \frac{\mathcal{C}\biggl[
		-\frac{\vec{h}\cdot\kT}{M_\mr{p}} f_1 D_1
		- \frac{(\vec{h}\cdot\pT)\kTs^2}{M_\mr{p}^2 M_\mr{h}} h_1^\perp H_1^\perp
		+ ...
		\biggr]}
		{\mathcal{C}\bigl[ f_1 D_1\bigr]},
	\quad
	\A{UU}{\cos2\phiH} = \frac{\mathcal{C}\biggl[
		-\frac{2(\vec{h}\cdot\kT)(\vec{h}\cdot\pT) - \kT\cdot\pT}{zM_\mr{p}M_\mr{h}}
			h_1^\perp H_1^\perp \biggr]}
		{\mathcal{C}\bigl[ f_1 D_1\bigr]},
\end{equation}
where
\begin{equation}
	\mathcal{C} \bigl[ w \, f \, F \bigr]
	= x \sum_q e_q^2 \int \dif{^2\pT} \dif{^2\kT} \, \delta(\PhT-\pT-z\kT)
			w f^q(x,\kT^2) F^q(z,\pT^2)
	\quad \mr{and} \quad
	\vec{h} = \frac{\PhT}{\PhTs}.
\end{equation}
In the expression for $\A{UU}{\cos\phiH}$ we have omitted the terms that are zero in the Wandzura--Wilczek approximation~\cite{Bastami:2018xqd}. The first of the remaining terms, proportional to the unpolarised TMDs $f_1$ and $D_1$, was suggested long ago by Cahn~\cite{Cahn:1978se} and could be used, in addition to the $\PhTs^2$-distribution of the $\sigma_0$, as an independent method of measuring $\langle \kTs^2 \rangle$.
The second term and also the expression for $\A{UU}{\cos2\phiH}$ contain the Boer--Mulders TMD PDF $h_1^\perp$~\cite{Boer:1997nt} and the Collins TMD FF $H_1^\perp$~\cite{Collins:1992kk}.
The amplitude $\A{LU}{\sin\phiH}$, known as the beam-spin asymmetry, is related to the twist-three functions $e$ and $g^\perp$ and to terms that are zero in the Wandzura--Wilczek approximation.
\section{COMPASS Measurements of the Azimuthal Modulations}
The amplitudes were measured by COMPASS in the production of charged hadrons in the scattering of muons off a $^6$LiD (effectively isoscalar) target~\cite{Adolph:2014pwc}. The results have been provided as a function of $x$, $z$ and $\PhTs$ simultaneously (3D) and integrated over two of the variables (1D). Later, the contribution of hadrons coming from the decay of exclusively (mostly diffractively) produced vector mesons was found sizeable in certain kinematic regions and subtracted~\cite{Agarwala:2019kqu}. The $\PhTs^2$-distributions have been measured as well~\cite{COMPASS:2017mvk}.

In 2016 and 2017 COMPASS collected a large dataset with 160~GeV/$c$ alternating $\upmu^+$ and $\upmu^-$ beams and a liquid hydrogen target. The beam was longitudinally polarised with $\lambda_{\upmu^-} \approx 0.8$ and $\lambda_{\upmu^+}\approx -0.8$
The azimuthal modulations and the $\PhTs^2$-distributions have been extracted using about 11\% of the sample~\cite{Moretti:2021naj}. The $\PhTs^2$-distributions together with corresponding distributions from $\mr{e^+e^-}$ annihilation at BELLE have been recently utilised in a leading-order determination of $\langle \kTs^2 \rangle$~\cite{Martin:2022}.
SIDIS events have been selected by requiring $Q^2>1~(\mr{GeV}/c)^2$ and $W>5$~GeV/$c^2$. The $Q^2$ cut implies $x>0.003$ in COMPASS kinematics. To stay in a region with reliable event reconstruction and moderate acceptance corrections $x<0.13$, $0.2<y<0.9$ and $\theta_{\upgamma^*}<60$~mrad have been imposed in addition, where $\theta_{\upgamma^*}$ is the polar angle of the virtual photon with respect to the beam. Only the hadrons with $z>0.1$ and $\PhTs>0.1$~GeV/$c$ have been selected to ensure good resolution in~$\phiH$.

As already mentioned above, part of the hadrons come from the decay of exclusive vector mesons, mostly $\uprho^0 \rightarrow \uppi^+\uppi^-$ and $\upphi \rightarrow \mr{K}^+\mr{K}^-$. The exclusive production of the vector mesons is dominantly diffractive, meaning that it can not be treated in the parton model. The hadrons coming from these decays are concentrated at low $Q^2$ (and thus low $x$) and low $\PhTs$. They populate quite uniformly the full $z$ range, while the rest of the hadron sample has a steeply decreasing $z$ distribution. As a result, their contribution is important at high $z$. As the vector mesons inherit the polarisation of the $\upgamma^*$, the decay hadrons obtain large azimuthal modulations, including $\cos\phiH$ and $\cos2\phiH$. To exclude this contribution, the events with only $\upmu^\prime \mr{h^+h^-}$ in the final state and having $z_\mr{h^+} + z_\mr{h^-} > 0.95$ have been rejected. The remaining contribution of partially reconstructed pairs has been determined from the HEPGEN Monte Carlo (MC)~\cite{Sandacz:2012at}, normalised to the data using the missing energy distribution of the reconstructed pairs, and subtracted in each $\phiH$ bin. The contamination fraction before the described treatment is shown in Fig.~\ref{fig:cont}. The acceptance correction has been determined utilising the LEPTO MC~\cite{Ingelman:1996mq}. No QED radiative corrections have been applied to these results, although work to determine them using the DJANGO MC~\cite{Charchula:1994kf} is ongoing.

\section{Preliminary Results on the Proton Target}

The azimuthal modulation amplitudes have been measured fitting the $\phiH$ distributions in bins of $x$, $z$ and $\PhTs$ in both the 1D approach (integrating over two variables) and 3D approach. In addition, the 1D results have been divided into four $Q^2$ ranges. The results obtained with the $\upmu^+$ and $\upmu^-$ beam have been found to be compatible and they have been merged. The uncertainties shown in the figures are only statistical. The systematic uncertainties have been estimated to be about the same size.

\begin{figure}[tbh]
\begin{minipage}[b]{0.675\textwidth}
\includegraphics[width=\textwidth]{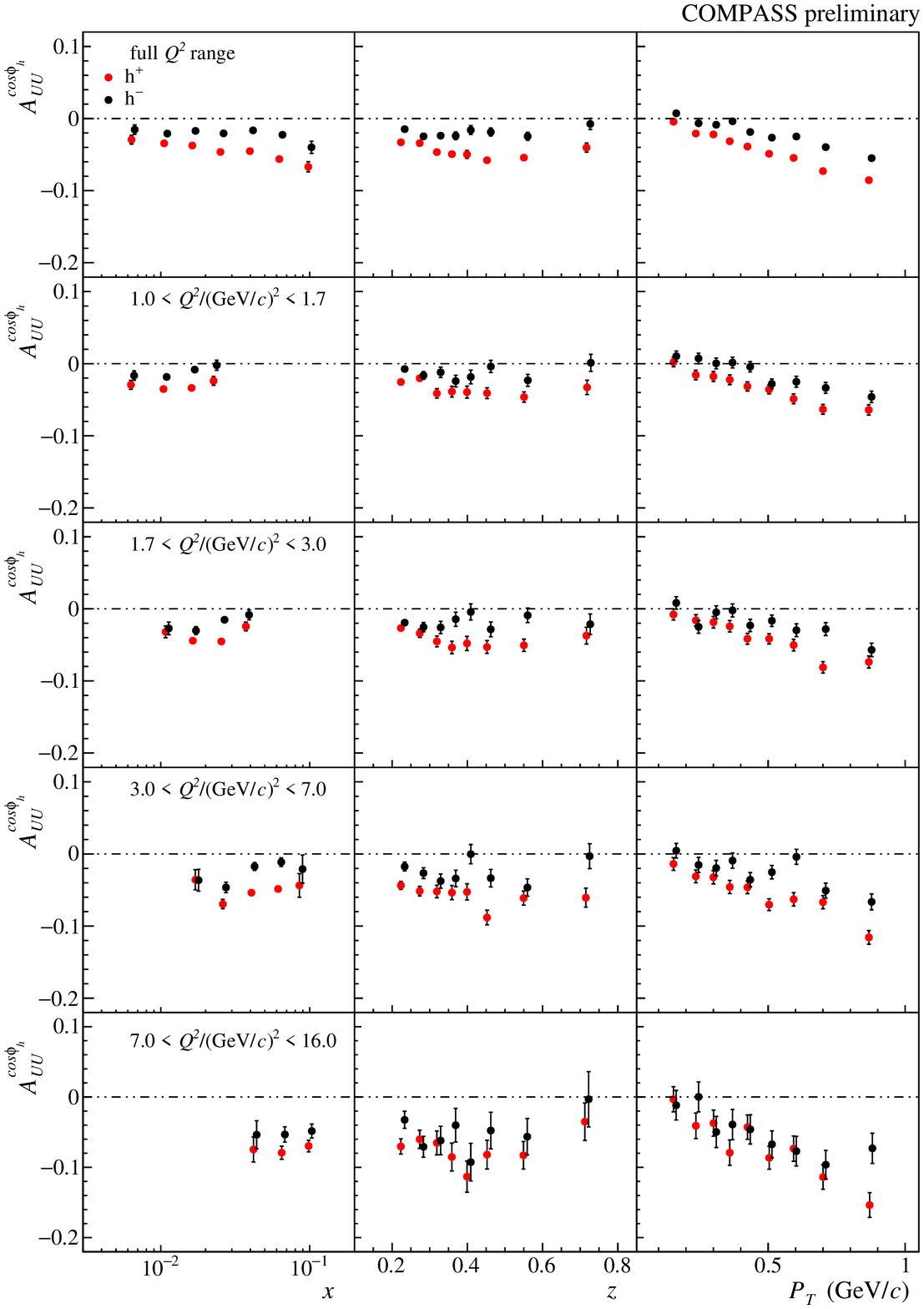}
\caption{\label{fig:acosphi_Q2comp}
	The amplitudes of the $\cos\phiH$ modulation in bins of $x$, $z$ and $\PhTs$
	in the full $Q^2$ range (top row) and divided in four $Q^2$ ranges.}
\end{minipage}
\hspace{0.01\textwidth}
\begin{minipage}[b]{0.305\textwidth}
    \includegraphics[width=0.8\textwidth]{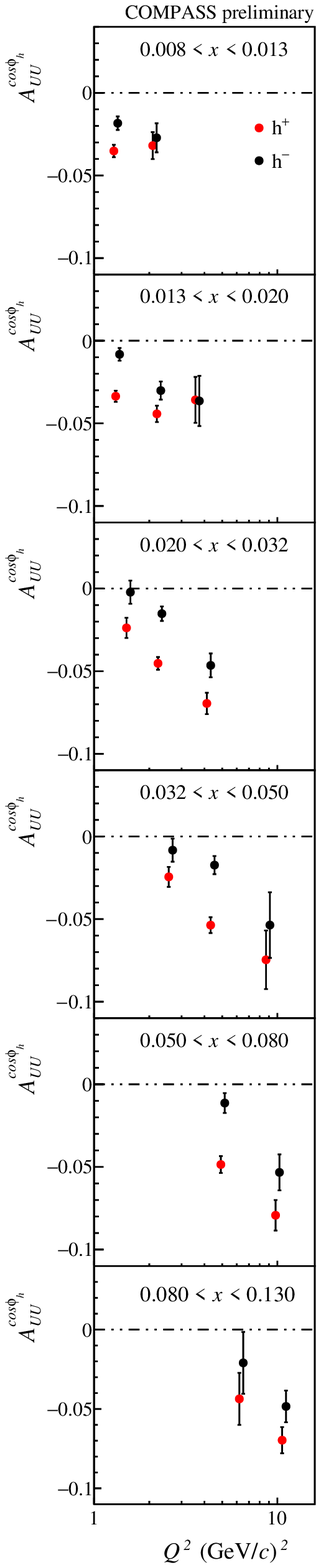}
    \caption{\label{fig:acosphi_Q2bins}
	The same amplitudes in bins of $Q^2$ in six $x$ ranges.}
\end{minipage}
\end{figure}

The 1D results for the amplitudes $\A{UU}{\cos\phiH}$ in the full $Q^2$ range are shown in Fig.~\ref{fig:acosphi_Q2comp}, top row. They are clearly non-zero and they differ between h$^+$ and h$^-$. The difference could point to a flavour-dependence of the $\langle \kTs^2 \rangle$. This can be further investigated comparing with the isoscalar target results~\cite{Adolph:2014pwc}, which exhibit a smaller difference. The results in bins of $(x,Q^2)$, $(z,Q^2)$ and $(\PhTs,Q^2)$ are shown in the remaining rows. Interestingly, the amplitudes increase with $Q^2$. This is counter-intuitive, as the Cahn effect, which is suppressed by $1/Q$, is expected to be the dominant contribution for this amplitude in Eq.~(\ref{eq:tmds}). It is worth reminding that $x$ and $Q^2$ are highly correlated in fixed-target kinematics and that the amplitudes increase with $x$. However, the increase of the amplitude with $Q^2$ is observed for six different ranges in $x$, as shown in Fig.~\ref{fig:acosphi_Q2bins}. Unfortunately, comparing this observation with the isoscalar target would require re-analysing the data as the $Q^2$ dependence was not extracted.

\begin{figure}[tbh]
\begin{minipage}[b]{0.675\textwidth}
\includegraphics[width=\textwidth]{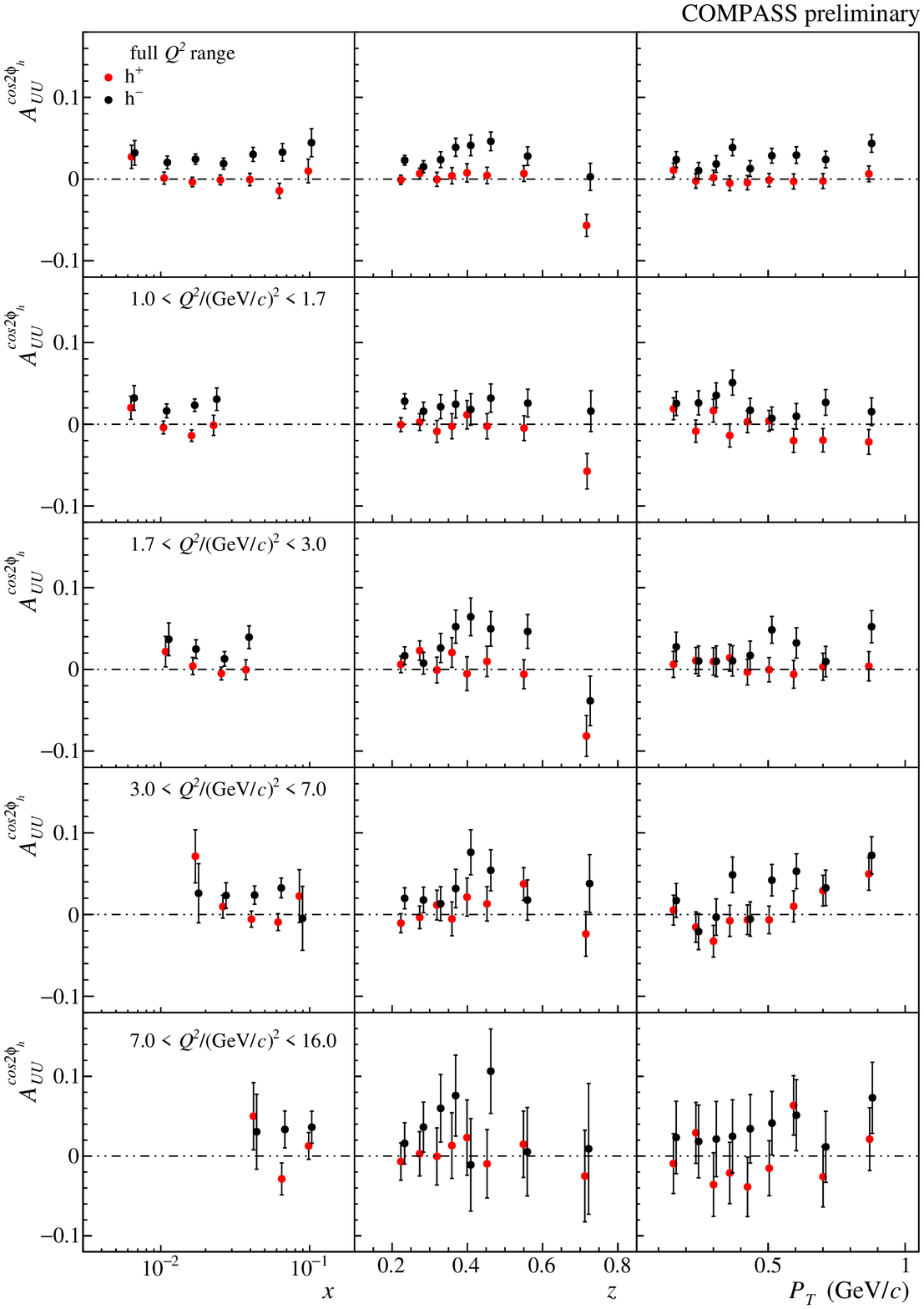}
\caption{\label{fig:acos2phi_Q2comp}
	The amplitudes of the $\cos2\phiH$ modulation in bins of $x$, $z$ and $\PhTs$
	in the full $Q^2$ range (top row) and divided in four $Q^2$ ranges.}
\end{minipage}
\hspace{0.01\textwidth}
\begin{minipage}[b]{0.305\textwidth}
    \includegraphics[width=0.8\textwidth]{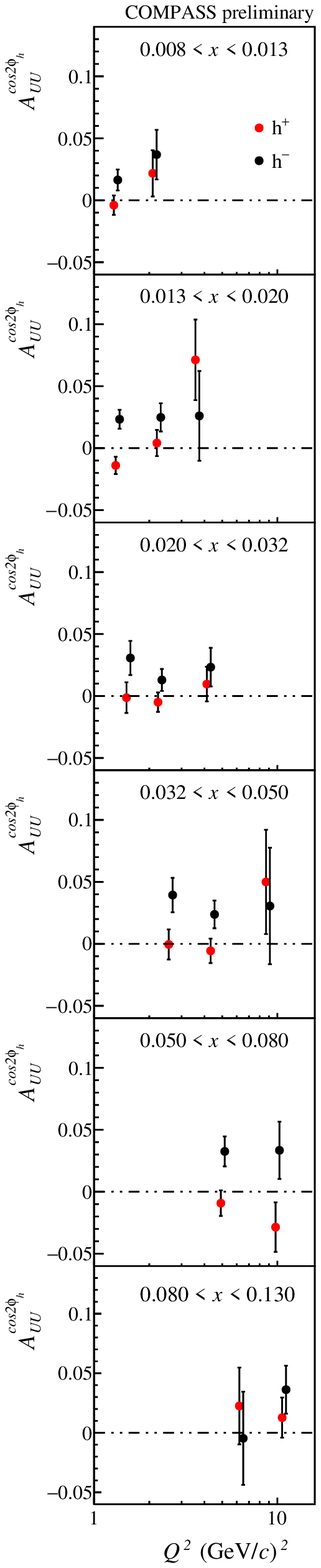}
    \caption{\label{fig:acos2phi_Q2bins}
	The same amplitudes in bins of $Q^2$ in six $x$ ranges.}
\end{minipage}
\end{figure}

\begin{figure}[tbh]
\begin{center}
\includegraphics[width=0.7\textwidth]{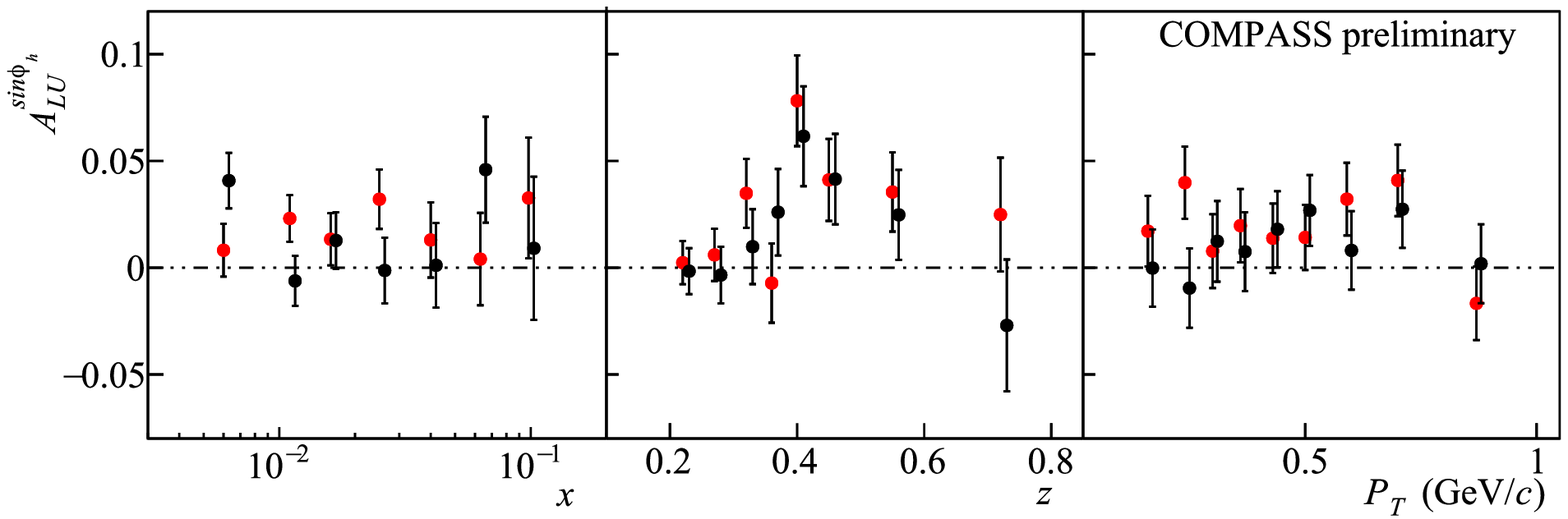}
\end{center}
\caption{\label{fig:asinphi}
	The amplitudes of the $\sin\phiH$ modulation in bins of $x$, $z$ and $\PhTs$}
\end{figure}

In the same manner, the amplitudes $\A{UU}{\cos2\phiH}$ are drawn in Fig.~\ref{fig:acos2phi_Q2comp},~\ref{fig:acos2phi_Q2bins}. The amplitudes for h$^+$ are generally compatible with zero, while those for h$^-$ are positive. No dependence on $Q^2$ is visible. The amplitudes $\A{LU}{\sin\phiH}$ in 1D are shown in Fig.~\ref{fig:asinphi}. The results are positive and compatible for h$^+$ and h$^-$.

\section{Acknowledgements}
The work of the author has been supported by the Czech MEYS grants LTT17018, LM2018104.
\providecommand{\href}[2]{#2}\begingroup\raggedright\endgroup
\end{document}